\documentclass[10pt,journal,a4paper]{IEEEtran}

\usepackage{amssymb}
\usepackage{amsmath}
\usepackage{graphicx}
\usepackage{subfigure}
\usepackage{cite}
\usepackage{enumerate}
\usepackage{bm}

\begin{document}

\title{Listen-and-Talk: Full-duplex Cognitive Radio Networks}

\author{
\IEEEauthorblockN{\small{Yun Liao}\IEEEauthorrefmark{1}, \small{Tianyu Wang}\IEEEauthorrefmark{1}, \small{Lingyang Song}\IEEEauthorrefmark{1}, \small{and Zhu Han}\IEEEauthorrefmark{2} \\}
\IEEEauthorblockA{\IEEEauthorrefmark{1}\small{State Key Laboratory of Advanced Optical Communication Systems and Networks,\\School of Electrical Engineering and Computer Science, Peking University, Beijing, China,}\\
\IEEEauthorrefmark{2}\small{Electrical and Computer Engineering Department, University of Houston, Houston, TX, USA.}\\}
}

\maketitle

\thispagestyle{empty}
\pagestyle{empty}

\begin{abstract}

In traditional cognitive radio networks, secondary users (SUs) typically access the spectrum of primary users~(PUs) by a two-stage ``listen-before-talk''~(LBT) protocol, i.e., SUs sense the spectrum holes in the first stage before transmit in the second stage. In this paper, we propose a novel ``listen-and-talk''~(LAT) protocol with the help of the full-duplex~(FD) technique that allows SUs to simultaneously sense and access the vacant spectrum. Analysis of sensing performance and SU's throughput are given for the proposed LAT protocol. And we find that due to self-interference caused by FD, increasing transmitting power of SUs does not always benefit to SU's throughput, which implies the existence of a power-throughput tradeoff. Besides, though the LAT protocol suffers from self-interference, it allows longer transmission time, while the performance of the traditional LBT protocol is limited by channel spatial correction and relatively shorter transmission period. To this end, we also present an adaptive scheme to improve SUs' throughput by switching between the LAT and LBT protocols. Numerical results are provided to verify the proposed methods and the theoretical results.

\end{abstract}

\IEEEpeerreviewmaketitle

\section{Introduction}%

With the fast development of wireless communication, spectrum resource has become increasingly scarce. Cognitive radio, as a promising solution to spectrum shortage, has caused wide attention for more than a decade\cite{mitola1999cognitive}\cite{mitola2000cognitive}. In cognitive radio networks~(CRNs), unlicensed or secondary users~(SUs) are allowed to opportunistically utilize the vacant slots in the spectrum allocated to primary users~(PUs). SUs therefore need to search for spectrum holes reliably and efficiently to protect the PU networks as well as maximize their own throughput\cite{akyildiz2006next}.

Traditionally, the so-called ``listen-before-talk''~(LBT) strategy in which SUs sense the target channel before transmission has been extensively studied\cite{yucek2009survey}. Optimization of sensing and transmission duration has been discussed in \cite{liang2008sensing} and \cite{huang2008short}. This LBT strategy requires little infrastructure support and it proves to be effective. However, it still has several problems such as sacrifice of transmitting time and discontinuity of transmission even if the white space of spectrum is continuous. The major reason is that most current deployed radios for wireless communications are half duplex such that to dissipate the precious resources by either employing time-division or frequency-division.

A full-duplex system, where a node can send and receive at the same time and frequency resources, offers the potential to double the spectral efficiency. However, due to the close proximity of a given modem¡¯s transmit antennas to its receive antennas, strong self-interference introduced by its own transmission makes decoding process nearly impossible, which is the reason why realization of FD techniques has not be deemed possible until recently. The last several years witness the advent of interference reduction techniques \cite{kiessling2004mutual} that provide the possibility of introducing FD to wireless communications. A number of works have been done to measure the performance of FD techniques which show that under certain circumstances, using FD can achieve better spectral efficiency than traditional half-duplex systems\cite{jain2011practical}.

Motivated by the FD technique, in this paper, we propose a ``listen-and-talk''~(LAT) protocol -- simultaneously sensing the spectrum and transmitting data for CRNs. At each moment, one of the antennas at each SU senses the target spectrum band, and judges if the PU is busy or idle; the other antenna transmits data simultaneously or keeps silent on the basis of the sensing results. Energy detection under imperfect self-interference suppression~(SIS) is used as the spectrum sensing strategy. We closely look into two cases when the SU is transmitting or remaining silent. Also, adaptive thresholds of detection are given.

To compare the LAT protocol with the conventional LBT protocol, we derive the probability of false alarm and miss detection, and the system throughput for these two protocols. We also analyze the power-throughput tradeoff for the LAT protocol. While overcoming discontinuity of transmission in conventional LBT protocol, the proposed LAT protocol still suffers the throughput loss by severe self-interference when transmit power increases. To this end, we propose an adaptive switching scheme between conventional and LAT approaches in pursuit of maximum throughput with the constraint of detection probability. Simulation results are provided to verify the proposed methods in terms of the signal-to-noise ratio~(SNR), average SU transmit power, SIS factor in the LAT protocol, spatial correlation coefficient and sensing duration in the conventional LBT protocol.

The rest of the paper is organized as follows. Section II describes the system models of the conventional LBT and the proposed LAT protocols. In Section III, we derive the analytical performance of the LAT protocol. In Section IV, we propose a switching scheme between the LBT and LAT protocols. Simulation results are presented to verify our analysis and visually show the throughput gain of the adaptive scheme, in comparison with both LBT and LAT protocols in Section V. We conclude the paper in Section VI.

\section{System Model}%

In this paper, we consider a CRN consisting of one PU and one SU pair, where SU$_1$ transmits data to SU$_2$. Each SU is equipped with two antennas Ant$_1$ and Ant$_2$. The spectrum band occupancy by the PU can be modeled as an alternating ON/OFF random process. The sensing and transmission process is time-slotted.

For simplicity and without loss of generality, energy detection is adopted as the sensing scheme, and the test statistics can be given as
\begin{equation}\label{test}
M = \frac{1}{{{N_s}}}\sum\limits_{n = 1}^{{N_s}} {{{\left| {y\left( n \right)} \right|}^2}},
\end{equation}
where $N_s$ denotes the number of samples, and $y(n)$ is the received signal of the $n^{th}$ sample. Let $f_s$ represent the sampling frequency in sensing process, and for sensing duration $t$, we have $N_s \propto f_st$.

We refer to the situation when PU is inactive as hypothesis $\mathcal{H}_0$, and the situation when PU is active is hypothesis $\mathcal{H}_1$. The probability of false alarm and miss detection can be written as
\begin{equation}\label{Pf}
{P_f}\left( \epsilon  \right) = \Pr \left( {M > \epsilon |{\mathcal{H}_0}} \right)
\end{equation}
and
\begin{equation}\label{Pd}
{P_m}\left( \epsilon  \right) = \Pr \left( {M < \epsilon |{\mathcal{H}_1}} \right),
\end{equation}
respectively, where $\epsilon$ denotes the detection threshold.

\vspace{-1em}
\subsection{Traditional Listen-before-Talk Protocol}

\begin{figure}[!t]
\includegraphics[width=3.6in]{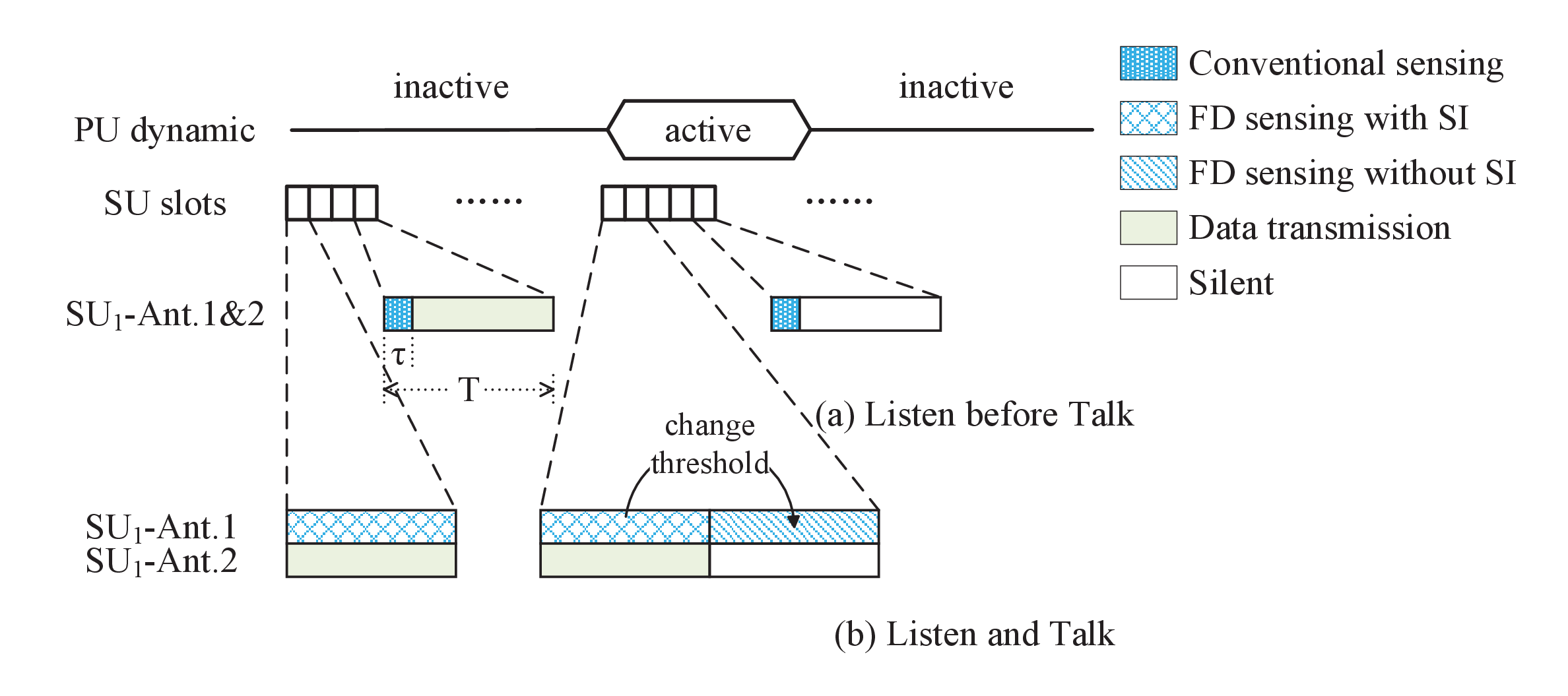}
\caption{System model: LBT and LAT protocols} \label{model}
\end{figure}

In this protocol, each slot $T$ is divided into two subslots as shown in Fig.~\ref{model}(a): sensing subslot with duration $\tau$, and data transmission subslot with duration $T-\tau$. Note that for comparison fairness with the FD based scheme, a 2$\times$2 MIMO is used for both spectrum sensing and data transmission. And the signal received at SU$_1$ in the sensing stage can be represented by
\begin{equation}\label{HDsense}
{\bf{y}} = \left\{
\begin{aligned}
& {{\bf{h}}_{\bf{s}}}s_p + {\bf{u}},& {\mathcal{H}_1},\\
& {\bf{u}},& {\mathcal{H}}_0,\\
\end{aligned}
\right.
\end{equation}
where $\bf{y}$, ${\bf{h}}_{\bf{s}}$, and $\bf{u}$ are $2\times1$ vectors representing the received signals, channels from PU to SU$_1$, and noise terms, respectively, and $s_P$ is the PU signal which is assumed to be PSK modulated with variance $\sigma_P^2$. The noise is independent and identically distributed~(i.i.d.) random Gaussian with zero mean and variance $\sigma _u^2$. With the separable correlation model\cite{kiessling2004mutual}, ${\bf{h}}_s$ can be expressed as ${{\bf{h}}_s} = {\bf{\Phi }}_s^{1/2}{{\bf{h}}_{s0}}$ where ${\bf{h}}_{s0}\sim \mathcal{CN}(0,\sigma _h^2)$, and ${\bf{\Phi }}_s$ stands for the normalized correlation matrix at SU$_1$.

The data are transmitted in a spatial multiplexing way, and at SU$_2$, we can have
\begin{equation}
{{\bf{r}}_2} = {{\bf{H}}_{12}}{{\bf{s}}_t} + {{\bf{u}}_2},
\end{equation}
where ${{\bf{H}}_{12}} = {\bf{\Phi }}_{r}^{1/2}{{\bf{H}}_{t0}}{\bf{\Phi }}_{t}^{1/2}$ is a $2 \times 2$ channel matrix from SU$_1$ to SU$_2$, $\bf{H}_{t0}$ is i.i.d. complex-valued Gaussian with zero mean and variance $\sigma_{\widetilde{h}}^2$, and ${{\bf{s}}_t}$ is the transmit signal vector with variance $\sigma_s^2$.

For simplicity, the exponential correlation model\cite{loyka2001capacity} is used for both sensing and transmission, and the correlation matrix can be represented by
\begin{equation}
{\bf{\Phi }} = \left( {\begin{array}{*{20}{c}}
1&{\beta}\\
{\beta^*}&1
\end{array}} \right),~\left|\beta\right| \in [0,1),
\end{equation}
where the correlation coefficient $\beta$ is the spatial correlation factor.

\vspace{-1em}
\subsection{Listen-and-Talk Protocol}

As shown in Fig.~\ref{model}(b), SU$_1$ performs sensing and transmission simultaneously using the FD technique: one of the two antennas at SU$_1$, say Ant$_1$, senses the spectrum while the other (Ant$_2$) transmits when a spectrum hole is detected. The challenge in this mode is that the transmit signal at Ant$_2$ of SU$_1$ is received by Ant$_1$, which causes self-interference at Ant$_1$. Thus, for sensing, the received signal is largely decided by the state of the other antenna: if Ant$_2$ of SU$_1$ is not transmitting, there is no difference to the conventional sensing method in \eqref{HDsense}, while if SU$_1$ is transmitting, self-interference will be introduced to the system. Given the differences above, we consider the circumstances when SU$_1$ is transmitting or not separately.

\subsubsection{Sensing without transmission}

In this case, the SU performs sensing by one antenna only, and we have
\begin{equation}\label{H0}
y = \left\{
\begin{aligned}
& h_ss_P + u,& {\mathcal{H}_{01}},\\
& u,& {\mathcal{H}}_{00},\\
\end{aligned}
\right.
\end{equation}
where $\mathcal{H}_{01}$ and $\mathcal{H}_{00}$ represent the hypothesises when SU$_1$ is silent, and the PU is busy or idle, respectively, and $h_s$ is the Rayleigh channel from PU to Ant$_1$ of SU$_1$, $u\sim \mathcal{CN}\left(0,\sigma_u^2\right)$ denotes the noise.

\subsubsection{Sensing and transmission}

With self-interference, the received signal can be written as
\begin{equation}\label{H1}
y = \left\{
\begin{aligned}
& h_ss_P + h_i s_t + u,& {\mathcal{H}_{11}},\\
& h_i s_t + u,& {\mathcal{H}}_{10},\\
\end{aligned}
\right.
\end{equation}
where $\mathcal{H}_{11}$ and $\mathcal{H}_{10}$ are the hypothesises under which SU$_1$ is transmitting and the PU is either busy or idle, respectively. $s_t$ in \eqref{H1} denotes the transmit signal at Ant$_2$ of SU$_1$ and $h_i$ represents the self-interference channel from Ant$_2$ to Ant$_1$. According to \cite{jain2011practical}, $h_i s_t$ can be modeled as a Rayleigh distribution with zero mean and variance $\chi^2\sigma_s^2$, where $\chi$ represents the SIS factor.

In the LAT protocol, only one link is used to data transmission and the received signal at Ant$_2$ of SU$_2$ is
\begin{equation}
r_2 = h_{12}s_t + u,
\end{equation}
where ${h}_{12}\sim \mathcal{CN}(0,\sigma _{\widetilde{h}}^2)$ is the transmit channel from SU$_1$ to SU$_2$, and $u\sim \mathcal{CN}(0,\sigma_u^2)$ represents the AWGN noise.

\section{Listen-and-Talk Protocol Analysis}%

In this section, we mainly study the analytical performance of the LAT protocol, derive throughput under the constraint of a given miss detection probability, and study the power-throughput tradeoff.

\vspace{-1em}
\subsection{Probability of Miss Detection and False Alarm}
Note that the test statistics $M_{LAT}$ is not only determined by PU activity, but also by the SU transmitter's behavior. We consider four conditions when PU is active or not, and SU$_1$ is transmitting or not, separately.

According to \eqref{test}, for each condition, the test statistics can be written as
\begin{equation}
M_{LAT} = \frac{1}{{{N_{s,LAT}}}}\sum\limits_{n = 1}^{{N_{s,LAT}}} {{{\left| {y\left( n \right)} \right|}^2}},
\end{equation}
where $N_{s,LAT} = f_sT$ is the samples number in the LAT protocol and $y$ takes different forms in different conditions. Note that $y$ in each sample is i.i.d. and we assume $N_{s,LAT}$ is large enough. According to central limit theorem (CLT), the PDF of $M_{LAT}$ can be approximated by a Gaussian distribution.

The statistical properties and the description under each condition are given in Table.~\ref{FDGaussian}, where $\gamma_s = \frac{\sigma_h^2\sigma_P^2}{\sigma_u^2}$ in Table.~\ref{FDGaussian} denotes the SNR in sensing, and $\gamma_i = \frac{\chi^2\sigma _s^2}{\sigma _u^2}$ is the interference-to-noise ratio~(INR). Detailed derivation of the distribution
properties are provided in Appendix A.

\begin{table}[!t]
\renewcommand{\arraystretch}{1.5}
\caption{Properties of PDFs of LAT} \label{FDGaussian}
\centering
\begin{tabular}{|c|c|c|c|c|c|}
\hline
Hypothesis & PU & SU & $\mathbb{E} \left[ {{M_{LAT}}} \right]$ & ${\mathop{\rm var}} \left[ {M_{LAT}} \right]$\\ \hline
$\mathcal{H}_{00}$ & idle & silent & $\sigma_u^2$ & $\frac{\sigma _u^4}{f_sT}$\\ \hline
$\mathcal{H}_{01}$ & busy & silent & $\left( {1+\gamma _s} \right)\sigma _u^2$ & $\frac{\left(1+\gamma_s\right)^2\sigma_u^4}{f_sT}$ \\ \hline
$\mathcal{H}_{10}$ & idle & active & $\left(1+\gamma_i\right)\sigma_u^2$ & $\frac{\left(1+\gamma_i\right)^2\sigma_u^4}{f_sT}$ \\ \hline
$\mathcal{H}_{11}$ & busy & active & $\left(1+\gamma_s + \gamma_i\right)\sigma_u^2$ & $\frac{\left(1+\gamma_s+\gamma_i\right)^2\sigma_u^4}{f_sT}$ \\ \hline
\end{tabular}
\end{table}

Let $\epsilon_0$ be the detection threshold when SU$_1$ is silent ($\mathcal{H}_{00},~\mathcal{H}_{01}$). Recalling the received signal in \eqref{H0}, and using \eqref{Pf}, the probability of false alarm ($P_{f,LAT}^0$) can be written as
\begin{equation}\label{PfF0}
P_{f,LAT}^0\left( \epsilon_0  \right) = \mathcal{Q}\left( {\left( {\frac{\epsilon_0 }{{\sigma _u^2}} - 1} \right)\sqrt {{f_s}T} } \right),
\end{equation}
where $\mathcal{Q}(\cdot)$ is the complementary distribution function of the standard Gaussian distribution. Furthermore, the probability of miss detection can be obtained from \eqref{Pd} as
\begin{equation}\label{PdF0}
P_{m,LAT}^0\left( \epsilon_0  \right) = 1 - \mathcal{Q}\left( {\left( {\frac{{{\epsilon _0}}}{{\left( {1 + {\gamma _s}} \right)\sigma _u^2}} - 1} \right)\sqrt {{f_s}T} } \right).
\end{equation}

For a given probability of miss detection $P_m^0$, the sensing threshold $\epsilon_0$ is given by
\begin{equation}\label{FDthhold0}
\epsilon_0  = \left( {\frac{{{Q^{ - 1}}\left( {1 - {P_m^0}} \right)}}{{\sqrt {{f_s}T} }} + 1} \right)\left( {1 + {\gamma _s}} \right)\sigma _u^2.
\end{equation}
Substitute \eqref{FDthhold0} to \eqref{PfF0}, we have the analytical false alarm probability written as
\begin{equation}
P_{f,LAT}^0\left( {{P_m^0}} \right) = \mathcal{Q}\left( {{\mathcal{Q}^{ - 1}}\left( {{1 - P_m^0}} \right)\left( {1 + {\gamma _s}} \right) + {\gamma _s}\sqrt {{f_s}T} } \right).\nonumber
\end{equation}

Similarly, when SU$_1$ is transmitting ($\mathcal{H}_{10},~\mathcal{H}_{11}$), and the detection threshold is $\epsilon_1$, the miss detection probability ($P_{m,LAT}^1$) and the false alarm probability ($P_{f,LAT}^1$) are
\begin{equation}\label{PdF1}
P_{m,LAT}^1\left( \epsilon_1  \right) = 1 - \mathcal{Q}\left( {\left( {\frac{\epsilon_1 }{{\left( {1 + {\gamma _s} + {\gamma _i}} \right)\sigma _u^2}} - 1} \right)\sqrt {{f_s}T} } \right),
\end{equation}
and
\begin{equation}\label{PfF1}
P_{f,LAT}^1\left( \epsilon_1  \right) = \mathcal{Q}\left( {\left( {\frac{\epsilon_1 }{{\left( {1 + {\gamma _i}} \right)\sigma _u^2}} - 1} \right)\sqrt {{f_s}T} } \right),
\end{equation}
respectively. And for a fixed probability of miss detection $P_m^1$, the false alarm probability can be derived as
\begin{equation}
\begin{split}
&{P_{f,LAT}^1}\left( {{P_m^1}} \right)\\
&= \mathcal{Q}\left( {{\mathcal{Q}^{ - 1}}\left( {{1 - P_m^1}} \right)\left( {1 + \frac{{{\gamma _s}}}{{1 + {\gamma _i}}}} \right) + \frac{{{\gamma _s}}}{{1 + {\gamma _i}}}\sqrt {{f_s}T} } \right).\nonumber
\end{split}
\end{equation}

\begin{figure}[!t]
\center
\includegraphics[width=2.4in]{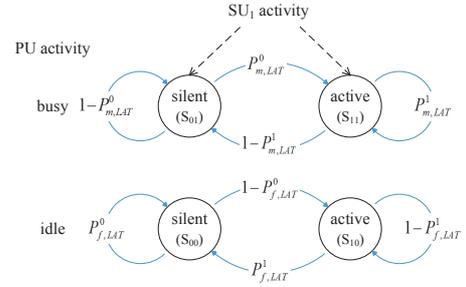}
\caption{SU state transition graph of the LAT protocol} \label{Markov}
\end{figure}

The changes among the four states are modeled as two discrete-time Markov chains (DTMC) illustrated in Fig.~\ref{Markov}, in which we assume that the PU's activity changes slowly compared to the duration of each time slot $T$, and a single slot when the PU changes its state can be neglected. Given that the probability for the system staying in each state $p_{ij}~(i,j=0,1)$ can be calculated considering the steady-state distribution of the Markov chains:
\begin{equation}
\left\{ {\begin{array}{*{20}{l}}
&{p_{01} \cdot P_{m,LAT}^0 = p_{11} \cdot \left(1 - P_{m,LAT}^1\right)},\\
&{p_{00} \cdot \left(1-P_{f,LAT}^0\right) = p_{10} \cdot P_{f,LAT}^1},\\
&{p_{01} + p_{11} = 1},\\
&{p_{00} + p_{10} = 1}.\\
\end{array}} \right.
\end{equation}

Then, the miss detection probability of the system is
\begin{equation}\label{PdF}
{P_{m,LAT}} = p_{11} = \frac{P_{m,LAT}^0}{1 + {P_{m,LAT}^0} - {P_{m,LAT}^1}},
\end{equation}
and the probability of the overall false alarm is
\begin{equation}\label{PfF}
{P_{f,LAT}} = p_{00} = \frac{P_{f,LAT}^1}{1 - {P_{f,LAT}^0} + {P_{f,LAT}^1}}.
\end{equation}
Within the limit of overall miss detection probability $P_{m,LAT} = P_{m}$, we obtain the constraints of $P_{m,LAT}^0$ and $P_{m,LAT}^1$ from \eqref{PdF}. For simplicity, we set $P_{m,LAT}^0=P_{m,LAT}^1$, and thus, $P_{m,LAT}^0=P_{m,LAT}^1=P_m$. From \eqref{PdF0} and \eqref{PdF1}, the test thresholds can be obtained according to SU$_1$'s activity as follow:
\begin{itemize}
\item When SU$_1$ is silent, using \eqref{PdF0}, we have
\begin{equation}\label{thhold0}
\epsilon_0\left(P_m\right) = \left( {\frac{\mathcal{Q}^{ - 1}\left(1 - P_m \right)}{\sqrt {{f_s}T} } + 1} \right)\left( {1 + {\gamma _s}} \right)\sigma _u^2.
\end{equation}
\item When SU$_1$ is active, using \eqref{PdF1}, the threshold $\epsilon_1$ is
\begin{equation}\label{thhold1}
\epsilon_1\left(P_m\right) = \left( {\frac{{{\mathcal{Q}^{ - 1}}\left( {{1 - P_m}} \right)}}{{\sqrt {{f_s}T} }} + 1} \right)\left( {1 + {\gamma _s} + {\gamma _i}} \right)\sigma _u^2.
\end{equation}
\end{itemize}

In shows in \eqref{thhold0} and \eqref{thhold1} that when SU$_1$ transmits, the detection threshold increases due to residual self-interference. Consequently, the probability of false alarm $P_{f,LAT}^0$ and $P_{f,LAT}^1$ are different:
\begin{equation}\label{PfFfinal}
\begin{split}
&P_{f,LAT}^0\left( {{P_m}} \right) = \mathcal{Q}\left( {{\mathcal{Q}^{ - 1}}\left( {{1 - P_m}} \right)\left( {1 + {\gamma _s}} \right) + {\gamma _s}\sqrt {{f_s}T} } \right),\\
&{P_{f,LAT}^1}\left( {{P_m}} \right)\\
&= \mathcal{Q}\left( {{\mathcal{Q}^{ - 1}}\left( {{1 - P_m}} \right)\left( {1 + \frac{{{\gamma _s}}}{{1 + {\gamma _i}}}} \right) + \frac{{{\gamma _s}}}{{1 + {\gamma _i}}}\sqrt {{f_s}T} } \right).
\end{split}
\end{equation}
Substituting \eqref{PfFfinal} to \eqref{PfF}, we can obtain the false alarm probability of the whole system.

\vspace{-1em}
\subsection{SU's Throughput}
During transmission, with transmit power $\sigma_s^2$, channel gain $h_t\sim \mathcal{CN}\left(0,\sigma_{\widetilde{h}}^2\right)$, and noise variance $\sigma_u^2$, the sum-rate can be written as
\begin{equation}
R_{LAT} = {\log _2}\left( {1 + \frac{{\sigma_s^2\sigma _{\widetilde h}^2}}{{\sigma _u^2}}} \right) = {\log _2}\left( 1+\gamma_t \right),
\end{equation}
where $\gamma_t$ represents the SNR in transmission. And the throughput can be expressed as
\begin{equation}\label{thputF}
C_{LAT} = R_{LAT} \cdot \left( 1 - P_{f,LAT} \right).
\end{equation}
It is shown in the expression of $R_{LAT}$ and $P_{f,LAT}$ that the throughput increases with SNR in sensing $\gamma_s$, and decreases with SIS factor $\chi$.

\subsection{Power-Throughput Tradeoff Analysis}

In the LAT protocol, $\sigma_s^2$ is positively proportional to $\gamma_t$, and thereby positively related to the sum rate. On the other hand, the power has strong influence on sensing performance, since it is also proportional to self-interference. Theoretically, with fixed SIS factor $\chi$, the sensing result deteriorates with the transmit power. With regards to the throughput, when transmit power is small, self-interference becomes negligible. The sensing results are reliable, and yet the throughput is limited. When the power is large, however, transmit power is no longer the limitation for the achievable sum rate ($R_{LAT}$), but self-interference may cause an unbearable high probability of false alarm. This may lead to severe waste of spectrum holes, which is also likely to decrease the throughput. Hence, there exists an optimal transmit power to achieve the best throughput. Note that due to space limitation, the mathematical proof will be given in our future work.

\section{Switching Between LAT and LBT}
There exist limitations for both LBT and LAT protocols. In the LBT, the data transmission time is reduced because of spectrum sensing, and the overall throughput is also affected by spatial correlation. In the LAT, residual self-interference is the main problem that decreases the performance. In this section, we first briefly derive the sensing performance and throughput in the LBT protocol, and then propose an adaptive switching scheme to maximize SU's throughput by selecting the right protocol between the LAT and LBT protocols for CRNs.

\vspace{-1em}
\subsection{Performance Analysis of LBT}
The test static $M_{LBT}$ can be generally written as
\begin{equation}
M_{LBT} = \frac{1}{N_{s,LBT}}\sum\limits_{n = 1}^{N_{s,LBT}} {\frac{{{\left| {y_1\left( n \right)} \right|}^2}+{{\left| {y_2\left( n \right)} \right|}^2}}{2}},
\end{equation}
where $N_{s,LBT} = f_s\tau$ is the number of samples in each sensing subslot, and $y$ is specified in \eqref{HDsense}. Again, the distribution of $M_{LBT}$ can be approximated by a Gaussian distribution according to CLT, given that each sample ${\frac{{{\left| {y_1} \right|}^2}+{{\left| {y_2} \right|}^2}}{2}}$ is i.i.d. and $N_{s,LBT}$ is sufficiently large.

The properties of the PDFs under both hypothesises are presented in Table.~\ref{HDGaussian}, in which $\beta_s$ denotes the spatial correlation coefficient in sensing. Detailed derivation of the distribution properties are provided in Appendix A.

\begin{table}[!t]
\renewcommand{\arraystretch}{1.5}
\caption{Properties of PDFs of the test statistics by LBT} \label{HDGaussian}
\centering
\begin{tabular}{|c|c|c|}
\hline
  & Mean ($\mathbb{E} \left[ {{M_{LBT}}} \right]$) & Variance (${\mathop{\rm var}} \left[ {M_{LBT}} \right]$)\\ \hline
$\mathcal{H}_0$ & $\sigma_u^2$ & $\frac{\sigma _u^4}{f_s\tau}$ \\ \hline
$\mathcal{H}_1$ & $\left( {\gamma _s + 1} \right)\sigma _u^2$ & $\frac{\left[ {{{\left( {{\beta _s}{\gamma _s}} \right)}^2} + {{\left( {{\gamma _s} + 1} \right)}^2}} \right]\sigma _u^4}{2f_s\tau}$ \\ \hline
\end{tabular}
\end{table}

The probabilities of false alarm and miss detection can be written, respectively, as
\begin{equation}\label{PH}
\begin{split}
{P_{f,LBT}}\left( {{\epsilon};\tau } \right)&= \mathcal{Q}\left( {\left( {\frac{{{\epsilon}}}{\sigma _u^2} - 1 } \right)\sqrt {{f_s}\tau } } \right),\\
P_{m,LBT}\left( {{\epsilon};\tau } \right) &= 1-\mathcal{Q}\left( {\frac{{\epsilon - \left( {\gamma _s + 1} \right)\sigma _u^2}}{{\xi \sigma _u^2}}\sqrt {{f_s}\tau } } \right),
\end{split}
\end{equation}
where $\xi := \frac{{{\left( {{\beta _s}{\gamma _s}} \right)}^2} + {{\left( {{\gamma _s} + 1} \right)}^2}}{2}$. And for a given probability of miss detection $P_m$, the analytical false alarm probability can be derived from \eqref{PH} as
\begin{equation}\label{PfH}
{P_{f,LBT}}\left( {{P_m};\tau } \right) = \mathcal{Q}\left( {{\mathcal{Q}^{ - 1}}\left( {{1-P_m}} \right)\xi  + \gamma _s\sqrt {{f_s}\tau }} \right).
\end{equation}

In transmission, with the constraint of average total power $\sigma_s^2$, the transmit power at each antenna is
\begin{equation}
P_{each} = \frac{{{\sigma_s^2}}}{2} \cdot \frac{T}{{T - \tau }},
\end{equation}
and the average sum rate is given by
\begin{equation}\label{HDrate}
{R_{LBT}} = \mathbb{E}\left[ {{{\log }_2}\det \left( {{\bf{I}} + \frac{{{P_{each}}}}{{\sigma _u^2}}{{\bf{H}}_{12}}{\bf{H}}_{12}^H} \right)} \right].
\end{equation}

At high SNR, $R_{LBT}$ in \eqref{HDrate} can be reduced as
\begin{equation}\label{approxrateH}
\begin{split}
{R_{con}}& \approx \mathbb{E}\left[ {{{\log }_2}\det \left( {\frac{{T\sigma_s^2}}{{2\left( {T - \tau } \right)\sigma _u^2}}{{\bf{H}}_{12}}{\bf{H}}_{12}^H} \right)} \right]\\
& = 2{\log _2}\left( {\frac{T}{{2\left( {T - \tau } \right)}}} \right) + 2\log _2 {\gamma_t} \\
&~~+ {\log _2}\left( {1 - \beta _t^2} \right) + {\log _2}\left( {1 - \beta _r^2} \right),
\end{split}
\end{equation}
where $\beta_t$ and $\beta_r$ represent the spatial correlation at SU transmitter (SU$_1$) and receiver (SU$_2$), respectively.

The throughput can be expressed as
\begin{equation}\label{thputH}
{C_{LBT}} = {R_{LBT}} \cdot \left( {1 - P_{f,LBT}} \right),
\end{equation}
which indicates that the throughput increases with transmit power $\sigma_s^2$ and SNR in sensing $\gamma_s$, and it decreases with the spatial correlation coefficients $\beta_s$, $\beta_r$, and $\beta_t$.

\vspace{-1em}
\subsection{Switching Algorithm}
Combining the throughput of conventional LBT and the proposed LAT protocols in \eqref{thputH} and \eqref{thputF}, respectively, the theoretical optimal switching criterion can be derived. Let $\Delta C$ be the difference of throughput between the two modes, we have
\begin{equation}
\Delta C = {C_{LBT}} - {C_{LAT}},
\end{equation}
and thus, the switching criterion is decided by the value of $\Delta C$:
\begin{equation}\label{switch}
{\text{operation~mode}} = \left\{
\begin{aligned}
&\text{Listen-before-talk},&\Delta C \ge 0,\\
&\text{Listen-and-talk},&\Delta C < 0.\\
\end{aligned}
\right.
\end{equation}

With $\Delta C = 0$, the optimal switching point can be easily calculated. Note that from \eqref{switch}, it implies that the switching point is related to the following statistical factors: SNR ($\gamma_s,~\gamma_t$) and transmit power ($\sigma_s^2$) during sensing and data transmission, spatial correlation coefficients ($\beta_s,~\beta_r,~\beta_t$) and the proportion of sensing time in a whole time slot ($\frac{\tau }{T}$) in the LBT protocol, and SIS factor ($\chi$) in the LAT protocol.

\section{Simulation Results}%

In this section, simulation results are presented to evaluate the performance of the proposed LAT protocol. Table.~\ref{simu-para} lists some important parameters in the simulation. For simplification, we set the spatial correlation coefficients $\beta_s = \beta_r = \beta_t = \beta$.

\begin{table}[!t]
\renewcommand{\arraystretch}{1.5}
\caption{Simulation Parameters} \label{simu-para}
\centering
\begin{tabular}{|l|l|}
\hline Parameters & Value \\ \hline
The duration of each time slot ($T$) & 0.2 ms \\ \hline
The duration of sensing time in LBT ($\tau$) & $0.25T$, $0.1T$ \\ \hline
The sampling frequency ($f_s$) & 1 MHz  \\ \hline
The number of samples in LAT ($N_{s,LAT}$) & 200  \\ \hline
The relative noise variance ($\sigma_u^2$) & 1  \\ \hline
The relative transmit power over noise ($P_t$) & $13$dB  \\ \hline
SNR in sensing process ($\gamma_s$) & $-10$dB  \\ \hline
SIS factor in LAT ($\chi$) & 0.4, 0.2  \\ \hline
The spatial correlation coefficient ($\beta$) & 0.7, 0.8 0.9  \\ \hline
Probability of miss detection ($P_m$) & 0.3 \\ \hline
\end{tabular}
\end{table}

\begin{figure}[!t]
\centering
\includegraphics[width=3.6in]{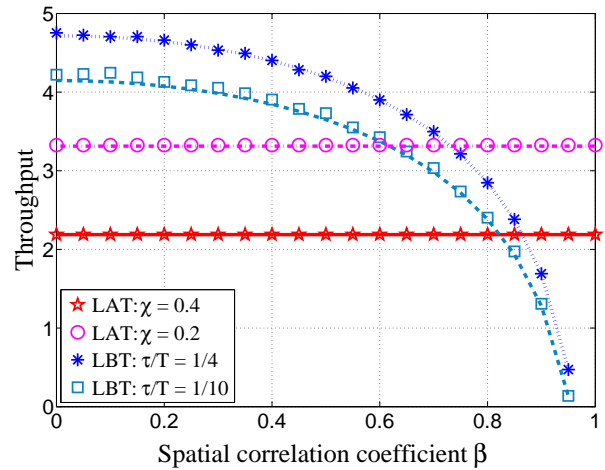}
\caption{SU's achievable throughput versus spatial correlation coefficient $\beta$} \label{beta}
\end{figure}

In Fig.~\ref{beta}, we consider the optimal switching point based on the spatial correlation coefficient, in which the probability of miss detection $P_m$ is fixed by 0.3, the sensing SNR is $-10$dB, and the relative transmit power is $13$dB. We investigate the cases when the SIS factor $\chi$ is 0.2 and 0.4, and when the sensing duration in the LBT protocol changes between $\frac{1}{4}$ and $\frac{1}{10}$. Fig.~\ref{beta} includes both analytical results (the real lines and dotted lines) and numerical results (various types of dots), which match perfectly. It can be shown that in the conventional LBT protocol, the achievable throughput decreases with the increment of spatial correlation, and to a certain point, the LAT protocol outperforms the conventional LBT protocol. Also, when residual self-interference increases, e.g., from 0.2 to 0.4, the performance of the LAT protocol becomes worse, and the switching point moves to a higher $\beta$.

\begin{figure}[!t]
\centering
\includegraphics[width=3.6in]{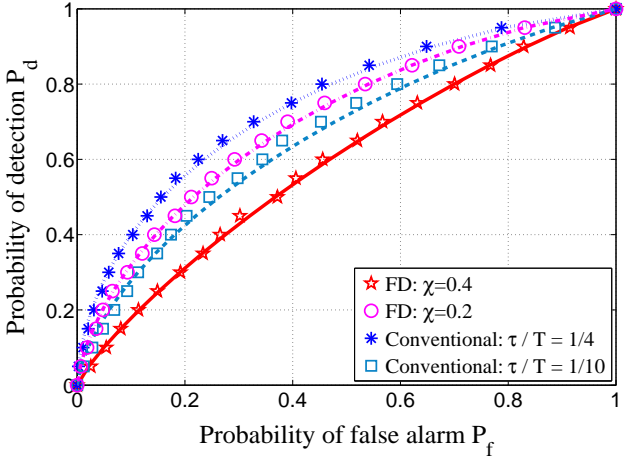}
\caption{ROC curve of both protocols} \label{ROCcurve}
\end{figure}

In Fig.~\ref{ROCcurve}, we use the receiver operating characteristic curve~(ROC) curves to present the sensing performance under different situations. With SIS factor $\chi = 0.2, 0.4$ in the LAT protocol, spatial correlation coefficient $\beta$ fixed on 0.7, and the sensing time takes up $\frac{1}{4}, \frac{1}{10}$ in each time slot in the conventional protocol, we have the relationship between the false alarm probability and miss detection probability. From Fig.~\ref{ROCcurve}, it is shown that the sensing performance becomes worse, i.e., $P_f$ increases and $P_m$ decreases with the increment of residual self-interference and the decrement of sensing time.

\begin{figure}[!t]
\centering
\includegraphics[width=3.6in]{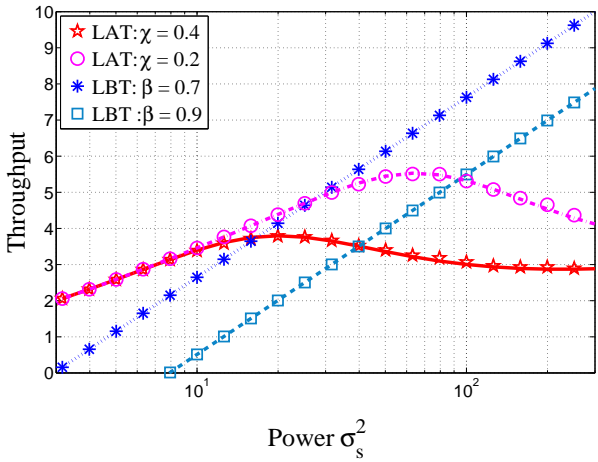}
\caption{Throughput of both protocols versus transmit power} \label{power}
\end{figure}

Fig.~\ref{power} evaluates the achievable throughput of SUs when transmit power changes within a certain range. We can observe that the power-throughput tradeoff in the LAT protocol, i.e., there exists an optimal transmit power in the low power range to achieve maximum throughput, and the optimal power decreases with the increment of SIS factor $\chi$. When the transmit power is low, due to longer transmit time and small residual self-interference, the LAT protocol can achieve better throughput. When transmit power becomes high, the LAT protocol suffers from severe self-interference while the conventional mode profits from the multiplexing gain, and thus, the conventional LBT protocol gradually becomes a better option.

\section{Conclusions}%

In this paper, we present a LAT protocol that allows SUs to simultaneously sense and access the spectrum holes. Besides, a switching scheme between the LAT and LBT protocols is provided to improve the throughput of SUs. Moreover, a tradeoff in LAT protocol between transmit power and the throughput is investigated by both analytical and numerical results. We find out that, the increment of transmit power does not always yield the improvement of SU's throughput, and a mediate value is required to achieve the best performance.

\appendices
\section{Proof of Table.~\ref{FDGaussian} and Table.~\ref{HDGaussian}}

We first provide the general properties of the test statistics. Given that each $y(n)$ in \eqref{test} is i.i.d., the mean and the variance of $M$ can be calculated as
\begin{equation}
\mathbb{E}\left[M\right] = \mathbb{E}\left[\left|y\right|^2\right];~{\rm{var}}\left[M\right] = \frac{1}{N_s}{\rm{var}}\left[\left|y\right|^2\right].\nonumber
\end{equation}
Further, if the received signal $y$ is complex-valued Gaussian with mean zero and variance $\sigma_y^2$, we have
\begin{equation}
\mathbb{E}\left[M\right] = \sigma_y^2,\nonumber
\end{equation}
and
\begin{equation}\label{CSCG}
{\rm{var}}\left[M\right] = \frac{1}{N_s}\left(\mathbb{E}\left[\left|y\right|^4\right]-\sigma_y^4\right)=\frac{\sigma_y^4}{N_s}.
\end{equation}

Then we consider the concrete form of the received signal under each hypothesis. In the LAT protocol, given the PU signal, residual self-interference, and i.i.d. noise, the received signal $y$ is complex-valued Gaussian with zero mean. The variance of $y$ under the four hypothesises are $\sigma_u^2~(\mathcal{H}_{00})$, $\left(1+\gamma_s\right)\sigma_u^2~(\mathcal{H}_{01})$, $\left(1+\gamma_i\right)\sigma_u^2~(\mathcal{H}_{10})$, and $\left(1+\gamma_s + \gamma_i\right)\sigma_u^2~(\mathcal{H}_{11})$, respectively. By substituting them into \eqref{CSCG}, we can obtain the results in Table.~\ref{FDGaussian}.

Distribution properties in the LBT protocol can be obtained by similar methods, and under hypothesis $\mathcal{H}_0$, we have $\sigma_y^2 = \sigma_u^2$. Under hypothesis $\mathcal{H}_1$, recalling \eqref{HDsense}, we have
\begin{equation}
\begin{split}
&\mathbb{E}\left[{\frac{{{\left| {y_1} \right|}^2}+{{\left| {y_2} \right|}^2}}{2}}\right]
=\frac{1}{2}\mathbb{E}\left[{\bf{y}}^H{\bf{y}}\right]\\
&=\frac{1}{2}\mathbb{E}\left[{\bf{h}}_{s0}^H({\bf{\Phi}}_s^{1/2})^H{\bf{\Phi}}_s^{1/2}{\bf{h}}_{s0}+2\sigma_u^2\right]=\left(\gamma_s+1\right)\sigma_u^2, \nonumber
\end{split}
\end{equation}
and the variance
\begin{equation}
\begin{split}
&{\rm{var}}\left[M\right] =\frac{1}{4N_s}\left(\mathbb{E}\left[{\bf{y}}^H{\bf{y}}\right]^2-4\left(1+\gamma_s\right)^2\sigma_u^4\right),~\text{in which}\\
&\mathbb{E}\left[{\bf{y}}^H{\bf{y}}\right]^2=\frac{1}{4N_s}\left(\mathbb{E}\left[\left|{\bf{h_s}}^H\bf{h_s}\right|^2+\left|{\bf{u}}^H{\bf{u}}\right|^2+4{\bf{h_s}}^H{\bf{h_s}}{\bf{u}}^H{\bf{u}}\right]\right)\\
&=\frac{1}{4N_s}\left(\left(2\left(3+\beta_s^2\right)\sigma_h^4+6\sigma_u^4+16\sigma_h^2\sigma_u^2\right)\right),\\
&\text{and thus,~}{\rm{var}}\left[M\right] =\frac{1}{2N_s}\left({\left[ {{{\left( {{\beta _s}{\gamma _s}} \right)}^2} + {{\left( {{\gamma _s} + 1} \right)}^2}} \right]\sigma _u^4}\right). \nonumber
\end{split}
\end{equation}

\bibliographystyle{IEEEtran}
\bibliography{IEEEabrv,REFS}

\begin{thebibliography}{20}
\bibitem{mitola1999cognitive}
J.~Mitola and G.~Q.~Maguire, ``Cognitive Radio: Making Software Radios more Personal," \emph{IEEE Personal Comm.}, vol.~6, no.~4, pp.13-18, Aug.~1999

\bibitem{mitola2000cognitive}
J.~Mitola, ``Cognitive Radio---An Integrated Agent Architecture for Software Defined Radio," Ph.D. Thesis, Royal Institute of Technology, Sweden, May.~2000.

\bibitem{akyildiz2006next}
I.~F.~Akyildiz, W.~Y.~Lee, M.~C.~Vuran, and S.Mohanty, ``Next Generation/Dynamic Spectrum Access/Cognitive Radio Wireless Networks: A Survey,'' \emph{Computer Networks}, vol.~50, no.~13, pp.~2127-2159, Sep.~2006.

\bibitem{yucek2009survey}
T.~Yucek and H.~Arslan, ``A Survey of Spectrum Sensing Algorithms for Cognitive Radio Applications,'' \emph{IEEE Comm. Surveys \& Tutorials}, vol.~11, no.~1, pp.~116-130, Mar.~2009.

\bibitem{liang2008sensing}
Y.~C.~Liang, Y.~Zeng, E.~C.~Y.~Peh, and A.~T.~Hoang, ``Sensing-Throughput Tradeoff for Cognitive Nadio Networks,'' \emph{IEEE Trans. Wireless Comm.}, vol.~7, no.~4, pp.~1326-1337, Apr.~2008.

\bibitem{huang2008short}
S.~Huang, X.~Liu, and Z.~Ding, ``Short Paper: On Optimal Sensing and Transmission Strategies for Dynamic Spectrum Access,'' in \emph{Proc. IEEE DySPAN}, Chicago, IL, Oct.~2008

\bibitem{kiessling2004mutual}
M.~Kiessling and J.~Speidel, ``Mutual Information of MIMO Channels in Correlated Rayleigh Fading Environments - a General Solution,'' in \emph{IEEE Int. Conf. Comm.}, vol.~2, pp.~814-818, Paris, France, Jun.~2004.

\bibitem{jain2011practical}
M.~Jain, J.~I.~Choi, T.~Kim, D.~Bharadia, S.~Seth, K.~Srinivasan, P.~Levis, S.~Katti, and P.~Sinha. ``Practical, Real-time, Full Duplex Wireless,'' in \emph{Proc. ACM MobiCom 2011}, New York, NY, Sep.~2011.

\bibitem{loyka2001capacity}
S.~L.~Loyka, ``Channel Capacity of MIMO Architecture Using the Exponential Correlation Matrix,'' \emph{IEEE Comm. Lett.}, vol.~5, no.~9, pp.~369-371, Sep.~2001.

\bibitem{huang2009optimal}
S.~Huang, X.~Liu, and Z.~Ding, ``Opportunistic Spectrum Access in Cognitive Radio Networks,'' in \emph{Proc. IEEE INFOCOM 2009}, Rio de Janeiro, Brazil, Apr.~2009




\end{thebibliography}

\end{document}